\documentclass[pagesize,12pt,a4paper]{scrartcl}
\usepackage{amsmath}
\usepackage{amsthm}
\usepackage{amssymb}
\usepackage{amsxtra}
\usepackage[squaren]{SIunits}
\usepackage{graphicx}
\usepackage{braket}
\usepackage{capt-of}
\usepackage{wasysym}
\usepackage{makeidx}
\usepackage{inputenc}
\usepackage[T1]{fontenc}
\usepackage{url}
\usepackage{eepic}
\usepackage{mathrsfs}
\usepackage{comment}
\usepackage{mparhack}
\def\out{\text{out}}
\def\inn{\text{in}}

\newcommand{\ir}{\mathrm{i}}
\newcommand{\e}{\mathrm{e}}

\newcommand{\eins}{{\mathbf 1}}

\renewcommand{\jmath}{j}

\newcommand*\idx[1]{\index{#1}#1}

\providecommand{\norm}[1]{\lVert#1\rVert}

\DeclareMathOperator{\dv}{d}

\DeclareMathOperator{\T}{T}

\DeclareMathOperator{\supp}{supp}
\DeclareMathOperator{\dt}{\tilde d\!}

\DeclareMathOperator*{\slim}{s-lim\,}

\ifx\KOMAScript\undefined%
  \DeclareRobustCommand{\KOMAScript}{\textsf{K\kern.05em O\kern.05em%
      M\kern.05em A\kern.1em-\kern.1em Script}}
\fi
\newlength{\help}
\newlength{\minuslaenge}
\newtheoremstyle{note}
  {3pt}
  {3pt}
  {\rmshape}
  {}
  {\bfseries}
  {:}
  {.5em}
  {}

\theoremstyle{note}

\makeatletter

 \def\vec#1{\ensuremath{\mathchoice
                     {\mbox{\boldmath$\displaystyle\mathbf{#1}$}}
                     {\mbox{\boldmath$\textstyle\mathbf{#1}$}}
                     {\mbox{\boldmath$\scriptstyle\mathbf{#1}$}}
                     {\mbox{\boldmath$\scriptscriptstyle\mathbf{#1}$}}}}%
\makeatother

\begin{document}

  \title{Relativistic Covariance of Scattering}
  \author{Norbert Dragon\\
          Institut f\"ur Theoretische Physik\\
          Leibniz Universit\"at Hannover 
}
\date{}

\maketitle

\begin{abstract} 

We analyze relativistic quantum scattering in the Schr\"odinger picture. The suggestive requirement
of translational invariance and conservation of the four-mo\-men\-tum, that the interacting Hamiltonian
commute with the four-momentum $P$ of free particles, is shown to imply the absence of interactions.

The relaxed requirement, that the interacting Hamiltonian $H'$ commute with the four-velocity $U= P/M$, $M=\sqrt{P^2}$,
allows Poincar\'e covariant interactions just as in the nonrelativistic case.
If the $S$-matrix is Lorentz invariant, it still commutes with the four-momentum $P$ though $H'$  does not.

Shifted observers, whose translations are generated by the four-velocity~$U$, just see a shifted superposition
of near-mass-degenerate states with unchanged relative phases, while the four-mo\-men\-tum generates 
oscillated superpositions with changed relative phases.

\end{abstract}

\newpage


\section{Introduction}

Despite the phenomenal agreement of the standard model with observed physics, 
the mathematical existence of relativistic scattering is still unknown.

In the quantum case and in the classical case relativistic scattering seems excluded by Haag's \cite{haag} and Leutwyler's \cite{leutwyler}  no-go theorems. 

In textbooks \cite{weinberg} one finds requirements for an interacting representation of Lorentz transformations 
but not their solution nor any proof of existence.

Assuming that one can switch on the interaction with a function $g:\mathbb R^4\mapsto [0,1]$, such that
$g(x)=0$ in a neighbourhood of $x$  means no interaction and $g(x)=1$ fully switched on interaction, 
and that the $S$-matrix is a series in $g$,  
Bogoliubov \cite{bogoliubov} shows that each unitary, perturbative, relativistic and causal $S$-matrix
is the time ordered exponential 
\begin{equation}
\label{svong}
S[g] =\T \exp{\,\ir\!\int\!\dv^{\,4}\!\! x \,\mathscr L_{\text{int}}(x, g(x))}\ ,
\end{equation}
where at each point $x$ the time ordered interaction Lagrangian $\T \mathscr L_{\text{int}}(x,g(x))$
is  a hermitian, scalar operator which depends on the intensity function $g$ and 
which is local 
\begin{equation}
\label{llocal}
\phantom{\,} [\T\mathscr L_{\text{int}}(x,g(x)),\T\mathscr L_{\text{int}}(y,g(y))] =0 \text{\qquad if $x-y$ is spacelike}\ .
\end{equation}
In the standard model $\T \mathscr L_{\text{int}}(x)$ is a normal ordered polynomial in the free fields at~$x$ 
which create and annihilate the elementary particles. 
But despite its fundamental role the convergence of the series is unknown as is the mathematical existence of relativistic scattering.

Mathematically well defined references on scattering theory \cite{reed3} restrict their discussion to nonrelativistic scattering
and specialize mainly to scattering by potentials which depend on the distance of the two incident particles.
Such an interaction is manifestly invariant under Galilei transformations, which govern nonrelativistic motion.

Using the Schr\"odinger picture we recapitulate the analysis of Reed and Simon in the relativistic case and find that the innocent looking
requirement of translational invariance, that the interacting Hamiltonian $H'$ commute with the generators $P^m$
of the translations of free particles, implies $H'=H$ and excludes scattering.

Each representation of the Poincar\'e group on many-particle states, however, is reducible and allows
the weaker invariance requirement that $H'$ commute with the four-velocity $U^m=P^m/M$, $M=\sqrt{P^2}$, 
which generates the translations of observers. Though $H'$ does not and must not commute with $P^m$, 
the resulting $S$-matrix does.

Using center coordinates we map relativistic scattering to the nonrelativistic case, thereby establishing its mathematical existence.
This shifts the requirement of locality (\ref{llocal}) into the focus of further investigations.

The usefulness of the Schr\"odinger picture is shown by the approximate factorization
of the scattering probability into the cross section and the integrated luminosity of the incident particles.
The latter is proportional to the spacetime overlap of the incident Schr\"odinger wavepacket and is basic to position 
measurements with light, which is plagued by the nonexistence of a position operator.

Notation: Let $T_{a}: x \mapsto x + a$ denote a translation in $\mathbb R^4$, $T_\Lambda: x \mapsto \Lambda x$ a Lorentz transformation, 
and $T_{a,\Lambda}=T_a T_\Lambda \in \mathfrak P$ a Poincar\'e transformation.
We denote by $U_{a,\Lambda}$ its unitary representation in a Hilbert space $\mathcal H_1$ of one-particle states.

\section{Free and Interacting Motion}

Two-particle states are spanned by products of one-particle states and naturally transform under
the Poincar\'e group by the product representation $U_{a,\Lambda}\otimes U_{a,\Lambda}$. 
Applied to two-particles states  $\Psi_2:(i,j,p_1,p_2)\mapsto \Psi^{ij}(p_1,p_2)$
the generators of translations, the momentum operators $P^m$,
satisfy the Leibniz rule and preserve the individual four-momenta, $p_1$ and $p_2$,
\emph{separately}, 
\begin{equation}
\label{p0free}
(P^m\, \Psi)^{ij}(p_1,p_2) = (p_1^m+p_2^m) \, \Psi^{ij}(p_1,p_2)\ .
\end{equation}
So the time evolution $\Psi(t)=\e^{-\ir H t}\Psi(0)$, generated by $H = P^0$, 
is \emph{free}.

An interacting time evolution must not map products of one particle states to the product of the freely evolved factors
but has to change the relative motion and the individual momenta of the scattering many-particle states. 

Let the Hamiltonian $H'$ generate the unitary one-parameter group of an interacting time evolution in the Hilbert space $\mathcal H$ of many-particle states
\begin{equation}
U'(t)=\e^{-\ir H' t}\ ,\ U'(t+s)=U'(t)U'(s)\ , \ t,s \in \mathbb R\ ,
\end{equation}
with worldlines $\Gamma=\set{(t, U'(t)\Psi), t\in \mathbb R}$ in quantum spacetime $\mathbb R \times \mathcal H$. 
At early times $U'(t)\Psi$ consists of distant particles, elementary or composite,  moving freely 
before they come near enough to interact.
The final state is considered sufficiently late such that the scattered particles
have separated, their mutual interactions have become
negligible and the particles move again freely. 

\begin{figure}[h]
\centering
\includegraphics[scale=.9]{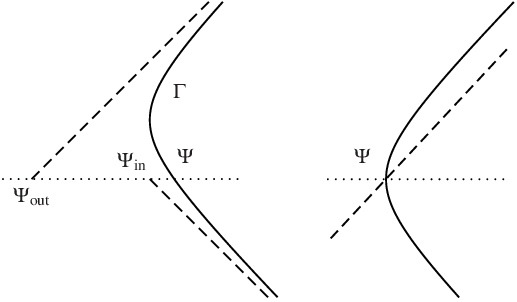}
\caption{Interacting path $\Gamma$ with asymptotes in $\mathbb R \times \mathcal H $, free and interacting path 
through~$\Psi$} 
\label{fig:scattering}
\end{figure}

To abstract from the inessential, one would like to consider the limits $\Psi_{\pm}$
of $U'(t)\Psi$ for $t\rightarrow \pm \infty$. 
\label{onepgroup}
But such limits of a unitary, nontrivial one-parameter group do not exist: 
$U'(s+t)=U'(s)U'(t)$ implies $\Psi_{\pm} =U'(s)\Psi_{\pm}$ and $U'(t)\Psi-\Psi_{\pm}=U'(t)(\Psi - \Psi_\pm)$.
As $U'(t)$ is unitary the norm of this difference  is time independent 
and vanishes in the limit only if it vanishes for all times.
i.e. only if $\Psi$ does \emph{not} move \cite{reed3}. 

A unitary, nontrivial group of motion has no limit.

The interacting evolution of scattering states $\Psi$ can at best approach the free evolution by $U(t)=\e^{-\ir H t}$
of asymptotic states $\Psi_\pm$ such that 
$U'(t)\Psi-U(t)\Psi_\pm$ or equivalently $U^{\prime -1}(t) U(t)\Psi_\pm $ converge for $t\rightarrow \pm \infty$. 
There have to exist the strong limits
\begin{equation}
\Omega_+ = \slim_{t\rightarrow \infty}\Omega(t)\ ,\ 
\Omega_- = \slim_{t\rightarrow -\infty}\Omega(t)\ ,\ 
\Omega(t) =\e^{\ir H'\,t}\,\e^{-\ir H\, t}\ ,\ 
\end{equation}
that the interacting path through each scattering state $\Psi$, which is orthogonal to all bound states of $H'$, 
has a past and a future free asymptote through states $\Psi_\inn$ and $\Psi_\out$ with
\begin{equation}
\label{waveop}
\lim_{t\rightarrow \infty }\Omega(t)\,\Psi_\out=  \Omega_+ \Psi_\out = \Psi\ ,\  
\lim_{t\rightarrow -\infty }\Omega(t)\,\Psi_\inn= \Omega_- \Psi_\inn =\Psi\ .
\end{equation}

The strong limit\index{strong limit}\index{$\slim$} of $\Omega(t)$ for $t\rightarrow \infty$ 
demands that for each $\varepsilon > 0$ and for each scattering state $\Psi$ 
there is a time~$T$ such that $\norm{(\Omega (t)-\Omega_+)\Psi}< \varepsilon \norm{\Psi}$ for all $t > T$.
This is a weaker condition than the uniform limit that $T$ be independent of~$\Psi$. 

To ask even stronger for the uniform limit of $\Omega(t)$ would require too much because at each time $T$ there are states 
which have not reached or left the interaction region. 

$\Omega_+$ and $\Omega_-$ are the generalized \idx{wave operators} or M\o ller operators.\index{Moller@M\o ller operator}

In more detail we write $\Omega_\pm(H',H)$ to display the involved Hamiltonians.  
One has  
$\Omega_{\pm}(H',H)^\star= \Omega_\pm(H,H')$ and
$\Omega_{\pm}(A,B)\,\Omega_{\pm}(B,C)=\Omega_{\pm}(A,C)$ \cite{reed3}.


By construction the wave operators $\Omega_{\pm}$ \emph{do not} commute with free time translations but intertwine
$H'$ unitarily with its corresponding $H$,
\begin{equation}
\label{moellerint}
\Omega_{\pm} = \slim_{t \rightarrow \pm\infty}\e^{\ir\, H'\,(t+a)}\e^{-\ir\, H\, (t+a)}= 
\e^{\ir \,H' a } \Omega_\pm \e^{-\ir\,H\,a}\ ,\quad
\e^{\ir \,H'\, a }\Omega_{\pm} = \Omega_{\pm} \e^{\ir \,H\, a} \ .
\end{equation}
Differentiation at $a=0$ shows
\begin{equation}
\label{scomH}
H' \Omega_\pm\Psi = \Omega_\pm H\Psi\ ,\ H'= \Omega_\pm H \Omega_\pm^{-1}\ ,
\end{equation}
for all smooth scattering states $\Psi$. On these states $H'$ is unitarily equivalent to $H$. 

But $H'$ must not commute with $H$, otherwise it commutes with $\Omega(t)$ and equals~$H$ if $\Omega_\pm$ exists. 
Thus, implementing translational invariance one must not require  $H'$ to commute with the unitary representation $U_a= \e^{\ir P a}$
of translations. 

\section{Cross Section and Luminosity}

We employ the Schr\"odinger picture and view time evolutions as worldlines $\set{(t,\Psi(t))}$ in quantum spacetime $\mathbb R \times \mathcal H$.
Interaction makes the worldlines of many-particle states depart from the free time evolution. 
$\Psi_\inn$ and $\Psi_\out$ are not states with momenta which are all directed towards or away from a scattering region. Rather 
they are the \emph{initial}  states of the future or past asymptotes which in the long run will automatically develop this property.
They only have to be many-particle states in the continuous spectrum of~$M'$.
That a state lies on an interacting trajectory is not a \emph{property} of the state but a \emph{relation} of the state and the path.
Such a relation does not contradict the additional relation that the same state also lies on a free trajectory.
By themselves states have no time evolution and are neither interacting nor free, they are just states
and determine the probabilities of the results of all measurements. 

Similarly in classical mechanics particles may traverse Kepler ellipses or straight lines, but this does not make the points of these curves elliptic
or straight. Nitpicking as the remark may seem, it spares the vain endeavours to construct interacting fields
or the futile considerations what an interacting Lorentz boost should be. Such denominations are widespread but misleading:
to be interacting is a property not of states but of time evolutions. Scattering theory  compares different time evolutions.

By the basic assumption of quantum theory the probability for a result $a_i$ (which for simplicity we take to be labeled
by some discrete index $i$) to occur if the state~$\Psi$ is measured with a perfect apparatus~$A$, is given by 
\begin{equation}
\label{qmbasic}
w(i,A,\Psi) = |\Braket{\Lambda_i | \Psi}|^2\ ,
\end{equation}
where $\Lambda_i$ is the state which yields $a_i$ with certainty.

In the \idx{Heisenberg picture} not the states evolve in the course of time $t$ but the operators which represent the measuring devices
and enter (\ref{qmbasic}) by their eigenvectors~$\Lambda$, 
\begin{equation}
\Braket{\Lambda(0)|\vphantom{U(t)^{\star}} U(t) \Psi(0)}_{\text{Schr\"odinger}}=\Braket{U(t)^{\star}\Lambda(0)|  \Psi(0)}_{\text{Heisenberg}}\ .
\end{equation}

The Heisenberg picture is invertibly related to the \idx{Schr\"odinger picture} and in this sense equivalent. 
But to ascribe the motion of \emph{several} particles, which move relative to each other and scatter, to the measuring devices
is as counterintuitive and misleading as the Ptolemaic system which describes the orbits of the planets 
in highly unsuitable, though admissible, coordinates in which the earth does not rotate  and does not orbit the sun. 
Try e.g. to understand the simple notion of the spacetime overlap of
colliding wave packets (\ref{lumin1}) in the Heisenberg picture. 


The Schr\"odinger picture does not rule out to consider time independent states, such as $\Psi_\inn$ or $\Psi_\out$, 
the initial states of the asymptotes of interacting paths,
nor does it preclude time dependent operators such as free fields. 
They are used to construct a local, relativistic $S$-matrix.
Time dependent fields do \emph{not} have to represent measuring devices in the Heisenberg picture,
notwithstanding axiomatic systems which call them \lq observables\rq .

The scattering matrix or $S$-matrix\index{Smatrix@$S$-matrix} is the map 
\begin{equation}
\label{sinout}
S:\Psi_\inn \mapsto  \Psi_\out\ ,\ 
S = \Omega_+^{\star}\Omega_-
= \slim_{t,\,t' \rightarrow \infty}\,\e^{\ir H t}\e^{-\ir H' (t+t')}\e^{\ir H t'}\ .
\end{equation}
Its matrix elements are scalar products of out- and in-states, 
$\braket{\Phi_\inn | \Psi_\out}=\braket{\Phi_\inn |S \Psi_\inn }$. 

By construction and by (\ref{moellerint}) the $S$-matrix commutes with temporal translations
\begin{equation}
\label{stemp}
S = \e^{\ir \,H\, a }\,S\,\e^{-\ir \,H\, a }\ ,\ [H, S] = 0\ .
\end{equation}
If the $S$-matrix commutes with Lorentztransformations, $U_\Lambda S\, U_\Lambda{}^{-1}=S$,
then $S$ commutes not only with $H=P^0$ but with all momenta $P^m$ and with all 
$U_{a,\Lambda}$. Conservation of $P^m$ does not require $H'$ to commute with $P^0$.

The relativistic $S$-matrix \emph{in the momentum basis}, not the basis independent $S$-matrix by itself, contains the experimentally 
available information about the interacting particles. 
Consider the transfer matrix $T := \ir(S - 1)$. Applied to an incoming two-particle state $\Psi$, omitting spin indices, 
using the short hand $q=(q_1,\dots q_n)$
and the scalar product of one-particle states on mass shells
$\mathcal M_m= \set{p:p^0 = \sqrt{m^2 + \vec p^2}, \vec p \in \mathbb R^3 }\subset \mathbb R^4$,
\begin{equation}
\braket{\Phi|\Psi}=\int\!\!\dt p\, \Phi(p)^*\, \Psi(p)\ ,\   
\dt p = \frac{\dv^3 p}{(2 \pi)^3\, 2 \sqrt{m^2 + \vec p^3}} \ ,
\end{equation}
its reduced kernel $\Braket {q | \mathfrak T | p_1,p_2}$ is defined by 
\begin{equation}
\label{tmatrix}
\ir (S-\eins)\Psi(q) = 
(2\pi)^4\!\! \int\! \tilde \dv p_1\, \tilde \dv p_2\,
\delta^4(p_1+p_2 - \sum_{i=1}^n q_i)
\Braket{q | \mathfrak T | p_1,p_2 }\Psi(p_1,p_2)\,.
\end{equation}
It determines the partial cross sections for the production of $n$ particles with momenta~$q$ in some domain $\Delta_n$ (denoting $\dt q_1\dots \dt q_n$ by $\dt^{\, n}\! q$)
\begin{equation} 
\label{sigmavons}
\sigma_{(p_1,p_2) \rightarrow \Delta_n}= \frac{(2\pi)^4}{4\sqrt{(p_1\cdot p_2)^2 - m_1^2m_2^2}}\int_{\Delta_n}\!\! \tilde \dv^n\! q\,
\delta^4(p_1+p_2 - \sum_{i=1}^n q_i)\, \bigl| \Braket{q | \mathfrak T | p_1,p_2 }\bigr|^2\ .
\end{equation}


We give a simple proof of this well-known basic relation of quantum scattering theory to observable physics,
which has the virtue to also determine the luminosity, which is basic to our optical perception of the world.

Recall that the wave function $\Phi=(S-1)\Psi$ is smooth if $\Psi$ is smooth: 
the relativistic $S$-matrix commutes with Poincar\'e transformations and maps the domain of the algebra of the generators,
rapidly decreasing smooth wave functions \cite{schmuedgen}, to itself. 
The $\delta^4$-function in (\ref{tmatrix}) is not a singularity but reduces the integral on $\mathcal M_1\times \mathcal M_2$ to the compact sub\-ma\-ni\-fold 
$ p_1 +  p_2 = \sum_{i=1}^n q_i$, the phase space of the reaction.
Similarly
\begin{equation}
\bigl((S^\star -1)\chi\bigr)(p_1,p_2)=\ir(2\pi)^4\!\! \int\! \tilde \dv^n q \,
\delta^4(p_1+p_2 - \sum_{i=1}^n q_i)\,
\Braket{ q| \mathfrak T |p_1,p_2  }^*\chi(q)
\end{equation}
is smooth if $\chi$ is smooth.

By the generalization of (\ref{qmbasic}) to results in a continuum, 
the integral $\int_{\Delta_n}\! \dt^n q\, \Phi^*(q)\,\Phi(q)$ is the probability to find after the scattering $n$ particles with momenta $q=(q_1,\dots q_n)$ 
in the domain $\Delta_n$. Inserting (\ref{tmatrix}) yields 4 momentum integrations with a product of $\delta^4$-functions of different variables
\begin{gather}
\nonumber
\delta^4(p_1 + p_2 - \sum_{i=1}^n q_i)\,\delta^4(p'_1 + p'_2 - \sum_{i=1}^n q_i)=\delta^4(p_1 + p_2 - \sum_{i=1}^n q_i)\,\delta^4(p'_1 + p'_2 - p_1 -p_2)\\
=\delta^4(p_1 + p_2 - \sum_{i=1}^n q_i)\,\frac{1}{(2\pi)^4}\int\!\dv^4 x\ \e^{\ir\, (p'_1 + p'_2 - p_1 -p_2)\,x}\ .
\end{gather}

So the second $\delta^4$-function can be exchanged by the spacetime integral over the products at $x$ of plane waves for each of the 
integration variables $(p_1,p_2,p'_1,p'_2)$. More precisely this applies if multiplied with smooth test functions, which is why we remarked 
that $\Psi$ and $(S-1)\Psi$ are smooth
and that the $\delta$-functions only reduce integrations to integrals over submanifolds.

In scattering of distinguishable particles the incoming state is a product of 
momentum wave packets,  $\Psi(p_1,p_2)=\Psi_1(p_1)\Psi_2(p_2)$, with support 
contained in small neighbourhoods around ${\bar p}_1$ and ${\bar p}_2$.
For small enough neighbourhood the smooth $\mathfrak T$-function does not vary appreciably. So we extract
\begin{equation}
\begin{gathered}
\delta^4(p_1+p_2 - \sum_{i=1}^n q_i)\,\Braket {q | \mathfrak T | p_1,p_2}\Braket {q | \mathfrak T | p'_1,p'_2}^*/\sqrt{p_1^0\,p_1^{\prime\,0}\,p_2^0\,p_2^{\prime\,0}}\\
\sim 
\delta^4(\bar p_1+\bar p_2 - \sum_{i=1}^n q_i)\, |\!\Braket {q | \mathfrak T | \bar p_1,\bar p_2}|^2/(\bar p_1^0 \bar p_2^0)
\end{gathered}
\end{equation}
as if constant from the $(p_1,p_2,p'_1,p'_2)$-integrations
of the wave packets. Each of these $p$-integrations is of the form 
\begin{equation}
\tilde \Psi(x) = \sqrt{2} \int\!\dt p\,\sqrt{p^0}\, \Psi(p)\,\e^{-\ir p\, x}
\end{equation}
or its complex conjugate and yields in the Schr\"odinger picture for massive particles the corresponding freely propagating position wave function at $x\in \mathbb R^{1,3}$, 
the remaining integration variable.  Dropping the symbol $\ \bar{}\ $  we find the momentum $q$ in the domain~$\Delta_n$ 
(which must not overlap with the beam) with probability
\begin{gather}
\label{wahrstreu}
w_{(p_1,p_2) \rightarrow \Delta_n}=\\
\nonumber
\frac{(2\pi)^4}{4 p^0_1 p^0_2 }\,\int_{\Delta_n}\! \dt^n q\, \delta^4(p_1 + p_2 - \sum_{i=1}^n q_i)\ |\Braket {q | \mathfrak T | p_1,p_2}|^2\,
\int\!\dv^4 x\, |\tilde \Psi_1(x)|^2\, |\tilde \Psi_2(x)|^2\ .
\end{gather}
It factorizes into the cross section (\ref{sigmavons}) times the integrated \idx{luminosity} $L$
\begin{equation}
\label{lumin1}
\begin{gathered}
w_{(p_1,p_2) \rightarrow \Delta_n}= \sigma_{(p_1,p_2) \rightarrow \Delta_n} \,L\ ,\\ 
L =  \frac{\sqrt{(p_1\cdot p_2)^2 - m_1^2m_2^2}}{p_1^0 p_2^0} \int\!\!\dv\!^{\,4}\! x\,  |\tilde \Psi_1(t,\vec x)|^2\,|\tilde \Psi_2(t,\vec x)|^2\ .
\end{gathered}
\end{equation}
Even if photons are massless and do not allow for a generator $\vec X$ of translations of spatial momentum, such that strictly speaking they do not have a well-defined
position wave function, we take the integrated luminosity (\ref{lumin1}) to define macroscopic position measurement: to detect an object you shine light on it and register the 
reflected light. The other way round: a beam of light becomes visible if traversing mist. 

During the overlap of wave packets we neglect their spreading which occurs because they are superposed of momenta 
near $p_1$ and $p_2$. Then in fixed target scattering the density $|\tilde \Psi_1|^2(t,\vec x)=\rho_1(\vec x)$ is time independent and
the impinging wave packet 
is rigidly shifted  with velocity $\vec v\ne 0$, $|\tilde \Psi_2|^2(t,\vec x)=\rho_2(\vec x- \vec v t)$. We employ coordinates  $\vec x =(x, x_\perp)$
parallel and perpendicular to $\vec v=(v,0,0)$ 
and denote the area densities obtained by integrating the volume densities along the beam by
\begin{equation}
\hat \rho(x_\perp)= \int\!\dv\! x\ \rho(x,x_\perp)\ .
\end{equation}
With these specifications the spacetime integral in (\ref{lumin1}) yields
\begin{equation}
\label{lumin}
\int\!\! \dv\!^{\,2}x_\perp\! \int\!\! \dv x\ \rho_1(x,x_\perp)\! \int\!\!\dv t\, \rho_2(x-vt,x_\perp)=
\int\!\! \dv\!^{\,2}x_\perp\, \hat \rho_1(x_\perp)\, \hat \rho_2(x_\perp)\, \frac{1}{v}\ .
\end{equation}

The beam overlaps the target, $\supp \hat \rho_1 \subset \supp \hat \rho_2$, else $\Psi_1$ is not a target in the beam. Moreover, within $\supp \rho_1$ 
the density of the beam $\hat \rho_2(x_\perp)=\bar \rho_2$ has its average value, at least after taking the mean of measurements with the target randomly positioned
in the beam. The remaining integral $\int\! \dv\!^{\,2}x_\perp\, \hat \rho_1(x_\perp)=1$ is the
number of targets. With $p_1=(m_1,0,0,0)$, $(p_1\cdot p_2)^2-m_1^2m_2^2= m_1^2 \vec p_2^2$ and $v= |\vec p_2| / p_2^0$ 
the integrated luminosity  for fixed target scattering turns out to be the mean area density of the beam, the inverse of the size $A$ of its transversal section,
\begin{equation}
L_{\text{fixed target}}=\bar \rho_2=1/A\ .
\end{equation}
The particle is scattered with the same probability $w=\sigma / A$
with which a randomly positioned point in the beam hits a fixed area 
of size $\sigma$ in the beam. This confirms that $\sigma_{(p_1,p_2)\rightarrow \Delta_n}$ (\ref{sigmavons}) 
is the partial cross section of the target.

\section{Center Variables}

The total momentum $P^m$ of an $n$-particle state $\Psi(p_1,\dots p_n)$ defines its invariant mass~$M$,
\begin{equation}
P^m\,\Psi(p_1,\dots p_n)=(\sum_{i=1}^n p^m_i)\Psi(p_1,\dots p_n)\ ,\ M^2 = P^2\ ,
\end{equation}
which has a purely continuous spectrum
\footnote{For $n\ge 2$ there are no eigenstates of $M$, as $P^2\Psi = m^2 \Psi$ restricts the support of $\Psi$
in the product of  mass shells \mbox{$\mathcal M_{1}\times \dots \times \mathcal M_n$} 
to a submanifold with vanishing $3n$-dimensional measure.
This continuous spectrum of~$M$ distinguishes many-particle states from one-particle states. 
}  with positive energies 
and allows to factorize $P^m$ as four-velocity $U^m$ times $M$
\begin{equation}
\label{ucenter}
P^m = U^m\,M\ ,\ U^2 = 1\ , \ [U^m, M]=0\ .
\end{equation}


To separate the motion of the center from the relative motion of the scattering particles, we change the variables of the wave function $\Psi$ from the momenta 
$(p_1,\dots p_n)$, $n\ge 2$, to the constrained center variables $(u,q)$ where 
\begin{equation}
u^m=\frac{\sum_i p^m_i}{\sqrt{(\sum_j p_j)^2}}\ ,\ u^2 = 1\ ,
\end{equation}
is the four-velocity of the center. It is well-defined unless all momenta are lightlike and colinear,
a Lorentz invariant sub\-mani\-fold $S^2\times \mathbb R^n$ which is outside the domain of scattering theory.
To obtain $q=(q_1, \dots q_{n})\in \mathbb R^{3n}$, the relative momenta at rest,  we decompose each momentum $p_i$ into parts which are parallel and orthogonal to~$u$,
\begin{equation}
p_i = p_{i\,\parallel} + p_{i\,\perp}\ ,\ p_{i\,\parallel} = (p_i\cdot u)\, u\ ,\ p_{i\,\perp} = p_i - p_{i\,\parallel}\ ,
\end{equation}
and boost each $p_{i\,\perp}$ by the inverse of the Lorentz boost
\begin{equation}
\label{lorp}
L_u = \begin{pmatrix}
\sqrt{1+\vec u^2}&  \vec  u^{\T}\\
 \vec   u &\quad  \eins+ \frac{ \vec  u\, \vec  u^{\T}}{1+\sqrt{1+\vec u^2}}\ ,
\end{pmatrix}
\end{equation}
which maps $\underline u= (1,0,0,0)$ to the four-velocity $u = (\sqrt{1+\vec u^2,} \vec u)$, to
\begin{equation}
q_i = (L_u)^{-1}p_{i\,\perp}\ .
\end{equation}
Its $0$-component vanishes, $0= u \cdot p_{i\,\perp} = (L_u^{-1} u)\cdot (L_u^{-1} p_{i\,\perp}) = \underline u \cdot q_i = q_i^0$: 
each \mbox{$q_i=(0,\vec q_i)$} lies in~$\mathbb R^3$. By definition,  $\sum_i p_i = \sum_i p_{i\parallel}$, so  the $q_i$
are constrained,
\begin{equation}
0=\sum_{i=1}^n q_i\ .
\end{equation}

Because of $m_i{}^2=(p_{i\,\parallel} + p_{i\,\perp})^2= (p_i\cdot u)^2 + q_{i}^2$, 
one has $(p_i\cdot u)^2=m_i^2 + \vec q_i^2 $ 
and the momenta $p_i$ in terms of the constrained center variables are
\begin{equation}
\label{cent}
p_i(u,q) = \sqrt{m_i^2 + \vec q_i^2}\,  u + L_u\,  q_i = \sqrt{m_i^2 + \vec q_i^2}
\begin{pmatrix}
\sqrt{1+\vec u^2}\\
\vec u
\end{pmatrix}
+
\begin{pmatrix}
\vec u \cdot \vec q_i\\
\vec q_i + \frac{(\vec u \cdot \vec q_i)\,\vec u}{1 + \sqrt{1 + \vec u^2}}
\end{pmatrix}\ .
\end{equation}
By $\sum_i p_i = \sum_i (p_i\cdot u)\, u = M \,u$ the invariant mass $M$ is the energy in the rest system
\begin{equation}
\label{M}
(M \Psi)(u,q) =M(q)\, \Psi(u,q)\ ,\  M(q)=\sum_{i=1}^n \sqrt{m_i^2 + \vec q_i^2}\ge \sum_i m_i\ .
\end{equation}

The momenta $p_i$ and 
$u$  Lorentz transform as four-vectors, $u \mapsto \Lambda u$, 
while for given $u$ the relative momenta~$q_i$ are Wigner rotated by 
$W(\Lambda, u)=L_{\Lambda u}^{-1} \Lambda L_u\in $SO$(3)$.

The constraint $\sum_i q_i = 0$ complicates $M(q)$. Solving it by $q_n=-\sum_{i=1}^{n-1} q_i$,
the mass $M(q)=\sum'\sqrt{m_i^2+ \vec q_i^2}+ \sqrt{m_n^2 +(\sum' \vec q_i)^2}$
depends for $n\ge 3$ not only on $\vec q_i^2$ but also on so called \idx{Hughes-Eckart term}s $\vec q_i\cdot \vec q_j$. 

Wavefunctions of the relative momenta $(q_1,\dots q_{n-1})$ together with the spins of the $n$ particles constitute a representation space of rotations SO$(3)$ or SU$(2)$. 
It decomposes into a sum $\sum_s \mathbb C^{2s+1}\otimes \mathcal I_s$ of multiplets on which the representation acts by 
multiplication with unitary spin-$s$ matrices with skew hermitian generators $\Gamma_{ij}$ leaving pointwise invariant the Hilbert space $\mathcal I_s$ of functions 
$f(r)$ of $d_n$ rotation invariant variables $r$ ($d_2=1$, $d_n=3(n-2)$ for $n > 2$). 
Each of these spin-$s$ multiplets of SO$(3)$ induces a representation $U_{a,\Lambda}$
of the Poincar\'e group $\mathfrak P$ in the space $\mathcal H_s \otimes \mathcal I_s $ of wave functions $\Psi(u,r)$
of the center's four-velocity~$u$, $u^2 = 1$, and of the invariants~$r$.  
The generators of $U_{a,\e^\omega}=\e^{\ir a P}\e^{-\ir \omega^{mn}M_{mn}/2}$, $M_{mn}= - M_{nm}$, act on these states by \cite{dragon}
\begin{gather}
\label{factoru}
P^m = U^m\, M\ ,\ \bigl(U^m \Psi\bigr)(u,r) = u^m\, \Psi(u,r)\ ,\  \bigl(M \Psi\bigr)(u,r) = M(r) \, \Psi(u,r)\ ,\\
\label{gencont}
\begin{aligned}
\bigl(-\ir M_{ij}\Psi\bigr)(u,r) &= -\bigl(u^i \partial_{u^j} - u^j \partial_{u^i}\bigr)\Psi(u,r) + \Gamma_{ij}\Psi(u,r)\ ,\\
\bigl(-\ir M_{0i}\Psi\bigr)(u,r) &= \sqrt{1+\vec u^2}\,\partial_{u^i}\Psi(u,r) + \Gamma_{ij}\frac{u^j}{1+\sqrt{1 +\vec u^2}}\Psi(u,r)\ .
\end{aligned}
\end{gather}
The states $\Psi$ are smooth not only as a function of $u$ but within open neighbourhoods also of the variables $r$, 
if they  are smooth functions of $(p_1\dots p_n)$ on each mass shell
as is required for~$\Psi$ to be 
in the domain of the generators which act by the product rule.

For a two-particle system and an observer at rest the Hamiltonian $ P^0=U^0\,M $  is
\begin{equation}
\label{qvonz}
H=\sqrt{1+\vec u^2}\bigl(\sqrt{m_1^2 + \vec q^2} + \sqrt{m_2^2 + \vec q^2}\bigr)=
\sqrt{1+\vec u^2}\bigl(m_1+m_2  + \frac{\vec z^2}{2\mu}\bigr)
\end{equation}
where $1/\mu = 1/m_1+1/m_2$. The 
function $\vec q^2(\vec z^2)$ exists by the implicit function theorem.
Explicitly it is given by 
\begin{equation}
\label{pvonm}
\vec{q}^{\, 2} = 
\frac{1}{4}(m^2 -2(m_1^2 + m_2^2) +  (m_1^2 - m_2^2)^2/m^2 )\ ,\ 
m^2(\vec z^2)=\bigl(m_1+m_2 + \vec z^2/(2\mu)\bigr)^2\ .
\end{equation}
To quadratic order in the velocities, $H$ is the nonrelativistic energy of the center of mass and of the relative motion 
confirming that the center variables generalize the center of mass coordinates to relativistic motion. However, the free relativistic Hamiltonian is
not the sum of the Hamiltonians of the center and the relative motion but their product. 

Let the states $\Psi$, which the standard observer~$\mathcal O$ measures with devices~$A$, 
be related by the unitary representation $V_{a,\Lambda}$ of $\mathfrak P$
to the states $\Psi_{a,\Lambda}$, which Poincar\'e transformed \idx{observer}s $\mathcal O_{a,\Lambda}$ measure with the same results  
with their devices $A_{a,\Lambda}$,
\begin{equation}
\label{obspoin}
A_{a,\Lambda}=V_{a,\Lambda}\, A\, V_{a,\Lambda}{}^{-1}\ ,\   \Psi_{a,\Lambda}=V_{a,\Lambda}\Psi\ ,\ 
w(i,A,\Psi)=w(i,A_{a,\Lambda}, \Psi_{a,\Lambda})\ .
\end{equation}

To satisfy the Poincar\'e algebra, the generators $\hat P^m = U^m\,\hat M$ of translations $V_{a,\eins}$ simply have to employ some hermitian, positive  $\hat M$ 
which commutes with the four-velocity $U^m$ and with the Lorentz generators $M_{mn}$.
However, only if $\hat M$ \index{$\mathfrak P, \mathcal O_{a,\Lambda}, T_{a,\Lambda}$} is a multiple of $1$, 
do the observers agree on all Poincar\'e invariant measurements with devices $1\otimes \hat A$, which only act on the invariant arguments $r$,
\begin{equation}
V_{a,\e^{\omega}}= \e^{\ir\, U\cdot a} \e^{-\ir \,\omega^{mn} M_{mn}/2}\ .
\end{equation}
This is the unique (up to the scale of $a$), maximal de\-ge\-ne\-rate representation of $\mathfrak P$ on many particle states.
These transformations $V_{a,\Lambda}$ correspond one-to-one to the observers~$\mathcal O_{a,\Lambda}$. As their translation is generated by
the four-velocity $U$, a superposition of nearly mass degenerate particles is seen by  translated observers
as the same superposition multiplied with a common, mass independent phase rather than an oscillated superposition with changed relative phases.



For the interaction $H'$ to be Poincar\'e covariant it has to commute with the translations $V_{a,\eins}$
of observers. Hence it commutes with $U^m$. 
And it has to Lorentz transform as $0$-component $H' = P^{\prime\,0}$ of a four-vector,
\begin{equation}
\label{comvu}
[U^m, H']=0\ ,\  V_{a,\Lambda}\,P^{\prime\,0}\,V_{a,\Lambda}{}^{-1} = \bigl( \Lambda^{-1}\bigr){}^0{}_m P^{\prime m}\ .
\end{equation}
But if $H'$ commutes with the hermitian, positive operator $U^0=\sqrt{1+\vec U^2}$ then 
both have a common spectral resolution and $M'=H'/U^0$ is a well-defined hermitian and Lorentz invariant operator,
\begin{equation}
\label{condition}
H'= U^0\, M'\ ,\ P^{\prime\, m}= U^m\, M'\ ,\ 
[M', U^m]=0\ ,\ [M', M_{mn}] = 0\ .
\end{equation}
Hence the M\o ller operators are Lorentz invariant,
\begin{equation}
\begin{gathered}
\lim_{t\rightarrow \pm \infty }\bigl(\e^{\ir\, u^0\,  M'\, t}\e^{-\ir\, u^0\,  M\, t}\Psi\bigr)(u,r)=
\lim_{t'\rightarrow \pm \infty }\bigl(\e^{\ir\, M' \,t'}\e^{-\ir\,  M\, t'}\Psi\bigr)(u,r)\ ,\\
\Omega_{\pm}(H',H)=\Omega_\pm(M',M)\ .
\end{gathered}
\end{equation}
So the $S$-matrix commutes with Lorentz transformations. By construction it commutes with $H=P^0$ (\ref{stemp}), 
hence it commutes with all $U_{a,\Lambda}$ and not only with $V_{a,\Lambda}$,
\begin{equation}
U_{a,\Lambda}\,S= S\,U_{a,\Lambda}\ .
\end{equation}

To commute with $U^m$ is a weaker restriction  of~$H'$ than to commute with $P^m$. The latter restriction excludes
scattering, as $\Omega(t) = \e^{\ir (H-H')t}$ has a limit only if $H=H'$. 

Therefore, in a basis of $\mathcal I_s$ in which~$M$ acts multiplicatively, $(Mf)(r)= m(r) f(r)$, the interacting mass $M'$ must not be multiplicative. 
For example $M'= M + V$ can be a sum with a potential $V(\vec x)$ where the position operator $\vec x$
constitutes Heisenberg pairs $[x^i,z^j]=\ir \delta^{ij}$ with the relative momentum $\vec z$ defined in (\ref{qvonz}).
If $V$ is spherically symmetric then the $S$-matrix commutes not only with rotations but also with boosts as they
act for each~$u$ by Wigner rotations of $\vec z$ and~$\vec x$. 

In nonrelativistic theory \cite{reed3} spherically symmetric potentials with a nontrivial $S$-matrix are known.
So also nontrivial relativistic scattering exists mathematically and not only as perturbation series with unknown convergence properties.
The relation of $M'$ and the time ordered interaction Lagrangian (\ref{llocal}) remains to be analyzed.

In the Hamiltonian description of \emph{classical} relativistic systems one can choose the generators of spatial translations and of rotations 
to coincide in the free and in the interacting case \cite{peres} 
\begin{equation}
\label{pint}
P^{\prime\,i}\stackrel{?}{=} P^i\ ,\  M'_{ij}=M_{ij}\ ,\ i,j\in \set{1,2,3}\ .
\end{equation}
In a quantum system these relations are suggested if one employs \lq pictures\rq\ for the interacting and free time evolution which coincide
at some time. As rotations and spatial translations are time independent and coincide at some time, their generators should agree at all times in the different pictures.
But these relations exclude interaction: Poincar\'e covariance requires the interacting Hamiltonian $H'$ to commute with the four-velocity $U^m$  (\ref{comvu}),
hence $P^{\prime\, m}= M'\,U^m$ and $P^{m}= M\,U^m$.  But then (\ref{pint}) implies $M'=M$ and $P^{\prime\,m}=P^m$ and
excludes scattering. 

Moreover, the condition (\ref{pint}) is measurably wrong as shown by the atomic weights of isotopes.
Their rate of spatial momentum transferred by their support to prevent free fall, their weight,  depends on the binding. 

Using (\ref{pint}) and exploiting ingeniously  the ana\-ly\-ti\-ci\-ty of Lorentz transformations, \index{Haag's theorem}
Rudolf Haag proved  \cite{haag} the absence of interaction and $S=1$ in axiomatic quantum field theory: 
the Wightman distributions, the vacuum expectation values of local fields, coincide at all times with the ones of free fields if the free and interacting fields coincide at an initial time.

Omitting all conditions the theorem is abbreviated to the statement that
the interaction picture only exists if there is no interaction. But it only excludes all theories based on the Wightman axioms.
They allow to reconstruct the fields and their transformations from the Wightman distributions up to unitary equi\-va\-len\-ce \cite{wightman}. 
But then, in case that there are no bound states, the distributions cannot distinguish the free from the interacting evolution as both differ only by unitary 
transformations~$\Omega_\pm$~(\ref{scomH}).

Haag's theorem is a no-go result of scattering theory as are the facts that a unitary group of motion has no limit, that $\Omega_\pm$  require the strong limit not the
norm limit and that on scattering states~$H$ and~$H'$ are unitarily equivalent but must not commute.

Haag's theorem is ignored by physicists who calculate successfully 
scattering amplitudes with Feynman graphs and Poincar\'e invariant rules to extract finite parts of products of free fields.
These calculations need no interacting fields. Strictly speaking 
$S = \T \exp\, \ir \int \dv^4\! x\, \mathcal L_{\text{int}}(x)$ is not a series in an algebra of operator valued distributions
as insinuated by the Wightman axioms.
Rather, the matrix elements of~$S$ are recursively  defined finite parts of integrals of products of distributions.

In Feynman graphs the term \lq interacting field\rq\  denotes an argument of
the time order~$\T$. But time order does \emph{not act on} operators but on a graded commutative algebra and \emph{yields} operator valued distributions.
The field's equation of motion \cite{zimmermann} in the abbreviated notation $\braket{\,X}:= \braket{\Omega | X \Omega }$
\begin{equation}
\label{varderop}
\braket{\,\T\,\e^{\ir \int\!\! \mathcal L_{\text{int}}}\,\frac{\hat \partial \mathcal L}{\hat \partial \phi}(x)\,\phi(x_1)\dots \phi(x_n)}
= \braket{\,\T\,\e^{\ir \int\!\! \mathcal L_{\text{int}}}\, \ir \frac{\delta }{\delta \phi(x)}\,\bigl(\phi(x_1)\dots \phi(x_n)\bigr)}
\end{equation}
contains the derivative ${\delta }/{\delta \phi(x)}$ confirming: an interacting field in a Feynman graph is an operation in a graded commutative algebra,
not an operator in Hilbert space.

\section{Position Measurement with Light}

Massless particles \emph{do not allow} a position operator $\vec X$, which generates translations of spatial momentum,
\begin{equation}
\label{transmom}
\bigl(\e^{\ir \vec b \cdot \vec X}\Psi\bigr)(\vec p)=\Psi(\vec p-\vec b)\ .
\end{equation}
It enlarges the algebra of the Poincar\'e generators by Heisenberg partners $X^j$ of the spatial momenta,
\begin{equation}
\ [P^i, P^j]=0=[X^i, X^j]\ ,\ [P^i, X^j] = -\ir\, \delta^{ij}\ ,  \ i,j\in\set{1,\dots D-1}\ . 
\end{equation}
Together with $P^0=\sqrt{\vec P^2}$ this algebra contains for $D > 2$
\begin{equation}
\sum_{j=1}^{D-1} [X^j,[X^j, P^0]]=-\frac{D-2}{|\vec P|}
\end{equation}
all powers of  $1/|\vec P|$. To be in the domain of this algebra, the wave functions have 
to decrease near $\vec p = 0$ faster than any power of $|\vec p|$. 

As the domain of the generators is invariant under the group which they generate \cite{schmuedgen} also all
$(\e^{\ir \, \vec b\,\vec X}\Psi)(\vec p)=\Psi(\vec p- \vec b)$ have to vanish at $\vec p=0$ for all~$\vec b$, 
thus $\Psi(\vec b)=0$ everywhere: 
the algebra of $\sqrt{\vec P^2}$, $\vec P$, the translations $\e^{\ir \vec b\cdot \vec X}$ of $\vec P$ and their generators $\vec X$
has no domain.



Different from massive particles the momentum spectrum of  massless particles contains a Lorentz fixed point, 
$p = 0$. There the function $p^0=\sqrt{\vec p^2}$ of $\mathbb R^{D-1}$  is only continuous but not smooth. 
This single, distinguished 
point is sufficient to spoil the translation invariance of spatial momentum. 
It prevents $P^0$ to enlarge the algebra of $\vec P$, the translations $\e^{\ir \vec b\cdot \vec X}$ and its generators~$\vec X$.
All attempts \cite{hawton, newton, pryce, wightman} to construct such generators for massless particles fail.

The position of a state  cannot be identified with the argument $x$
of the field $\Phi(x)$ which creates and annihilates the particle.  
By the Reeh-Schlieder theorem \cite{reeh} the operators
$\Phi_f = \int\! \dv^4 x \,f(x) \Phi(x) $ with support of $f$ contained in a fixed open set $\mathcal U$
create out of the vacuum a dense subspace of one-particle states. So their position cannot be restricted to $\mathcal U$.

That there is no position operator for massless particles disappoints expectations, because we see the world and reconstruct the position of all objects by light
which we receive as flow of massless quanta. But we do not see a distant photon. Rather we see massive objects,
using (\ref{lumin}),  by the 
currents of photons which they emit or scatter and which are annihilated in our retina.


\section{Conclusions}
The correct covariance requirement allows relativistic scattering. 
It differs from the requirements used in Haag's and Leutwyler's no-go theorems.

The factorization of the scattering probability into cross section times luminosity
holds only in the approximation that in momentum space the colliding wave packets are narrow
as compared to scales on which scattering amplitudes vary appreciably and are in addition localized 
in spacetime precisely enough to define their overlap. Neither the limit of sharp wave packets 
in momentum space nor in spacetime exist. It remains conceptually dubious what scattering in strongly curved spacetime is.

The Hamiltonians of the relative motion and the motion of the center do not decompose into a sum but are a product.
Bound states are eigenstates of the interacting invariant mass. Its relation to the local interaction 
Lagrangian still needs clarification.

\end{document}